\documentclass[%
 reprint,
superscriptaddress,
 amsmath,amssymb,
 prl,
]{revtex4-2}

\usepackage{titlesec}%
\usepackage{subfigure}%
\usepackage{amsfonts}%
\usepackage{graphicx}%
\usepackage{setspace}%
\usepackage{comment}%
\usepackage{dsfont}%
\usepackage{color}%
\usepackage{tikz}
\usetikzlibrary{arrows,calc}
\usepackage[colorlinks=true,linkcolor=blue,citecolor=blue,urlcolor=blue]{hyperref}

\newcommand{\ket}[1]{\left| #1 \right>} 
\newcommand{\bra}[1]{\left< #1 \right|} 

\titleformat{\paragraph}
{\normalfont\normalsize\bfseries}{\theparagraph}{1em}{}
\titlespacing*{\paragraph}
{0pt}{3.25ex plus 1ex minus .2ex}{1.5ex plus .2ex}

\begin{document}
\title{Unveiling higher-order topology via polarized topological charges}
\author{Wei Jia}
\affiliation{Key Laboratory of Quantum Theory and Applications of MoE, Lanzhou Center for Theoretical Physics, and Key Laboratory of Theoretical Physics of Gansu Province, Lanzhou University, Lanzhou 730000, China}
\author{Bao-Zong Wang}
\affiliation{International Center for Quantum Materials, School of Physics, Peking University, Beijing 100871, China}
\author{Ming-Jian Gao}
\affiliation{Key Laboratory of Quantum Theory and Applications of MoE, Lanzhou Center for Theoretical Physics, and Key Laboratory of Theoretical Physics of Gansu Province, Lanzhou University, Lanzhou 730000, China}
\author{Jun-Hong An}
\email{anjhong@lzu.edu.cn}
\affiliation{Key Laboratory of Quantum Theory and Applications of MoE, Lanzhou Center for Theoretical Physics, and Key Laboratory of Theoretical Physics of Gansu Province, Lanzhou University, Lanzhou 730000, China}
\begin{abstract}
Higher-order topological phases (HOTPs) host exotic topological states that go beyond the traditional bulk-boundary correspondence. Up to now, there is still a lack of experimentally measurable momentum-space topological characterization for the HOTPs, which is not conducive to revealing the essential properties of these topological states and also restricts their detection in quantum simulation systems. Here, we propose a concept of polarized topological charges to characterize chiral-symmetric HOTPs in momentum space, which further facilitates a feasible experimental scheme to detect the HOTPs in $^{87}$Rb cold atomic system. Remarkably, our characterization theory not only shows that the second-order (third-order) topological phases are determined by a quarter (negative eighth) of the total polarized topological charges, but also reveals that the higher-order topological phase transitions are identified by the creation or annihilation of polarized topological charges. Particularly, these polarized topological charges can be measured by pseudospin structures of the systems. Due to theoretical simplicity and observational intuitiveness, this work shall advance the broad studies of the HOTPs in both theory and experiment.
\end{abstract}
\maketitle

{\it\color{blue}Introduction.}---Topological phases of matter always attract the great attention in condensed matter physics. In recent years, higher-order topological phases (HOTPs) have aroused the broad interests because of showing the novel bulk-boundary correspondence~\cite{benalcazar2017quantized,langbehn2017reflection,song2017d,benalcazar2017electric,khalaf2018higher}, allowing the systems host the lower-dimensional topological edge states~\cite{liu2017novel,ezawa2018higher,schindler2018higher,khalaf2018symmetry,liu2019helical,ghorashi2020vortex,kheirkhah2020first,lichangan2020topological,wu2020boundary,ezawa2020edge,asaga2020boundary,claes2020wannier,li2020topological,yang2020type,khalaf2021boundary,ghosh2021hierarchy,mao2022orbital,chen2022experimental,liu2023analytic,yang2024higher,huang2024surface}, and inducing the potential applications in quantum computation~\cite{zhang2020topological} and quantum interferometer~\cite{li2021higher}. A variety of HOTPs have been discovered in insulators~\cite{trifunovic2019higher,queiroz2019splitting,park2019higher,sheng2019two,cualuguaru2019higher,tiwari2020unhinging,zhang2020mobius,chen2020higher,noguchi2021evidence,wei20213d,noguchi2021evidence,du2022acoustic,jia2023unified}, semimetals~\cite{lin2018topological,ahn2018band,wang2019higher,xu2020high,wang2020higher,ghorashi2020higher,liu2021higher,pu2023acoustic}, and superconductors~\cite{yan2018majorana,hsu2018majorana,yan2019higher,zhu2019second,volpez2019second,schindler2020pairing,tan2022two,chew2023higher,zhu2024direct}.  The studies have also been extended in Floquet~\cite{peng2019floquet,huang2020floquet,hu2020dynamical}, non-Hermitian~\cite{luo2019higher,lee2019hybrid,zhang2019non,liu2019second}, interacting~\cite{stepanenko2022higher,may2022interaction}, and fractal systems~\cite{yang2020photonic,zheng2022observation}. Nevertheless, there is still a fundamental issue that how to uniformly characterize and directly detect the chiral-symmetric HOTPs~\cite{altland1997nonstandard,kitaev2009periodic,ryu2010topological,chiu2016classification}, due to the sophisticated topological origins rooted in bulk or boundary of the systems.

The recent breakthrough has elucidated that real-space topological invariants~\cite{benalcazar2022chiral,lin2024probing,li2024exact} are able to capture the chiral-symmetric topological states protecting multipole zero modes at each corner, which enables the realization of chiral-symmetric HOTPs in real space by simulating boundary physics via the classical systems, such as acoustic crystals~\cite{wang2023realization,li2023acoustic} and electric circuits~\cite{li2023observation}. However, the bulk topologies of the HOTPs are associated with the global feature of Bloch wave functions in Brillouin zone (BZ)~\cite{hasan2010colloquium,qi2011topological}, which implies that the momentum-space topological invariants are essential reflection of their bulk topologies. Hence it is necessary to develop the momentum-space topological characterization theory to reveal the fundamental properties of these chiral-symmetric HOTPs. On the other hand, the proposed real-space topological invariants are not conducive to detect these HOTPs in quantum simulation systems with high controllability, such as ultracold atoms~\cite{bloch2012quantum,jotzu2014experimental,wu2016realization,schafer2020tools}, nitrogen-vacancy center~\cite{kong2016direct,ariyaratne2018nanoscale,ji2020quantum}, and nuclear magnetic resonance~\cite{xin2020experimental,zhao2021characterizing}, as the bulk physics defined in momentum space are conveniently simulated in these artificial systems. There is an urgent need an experimentally measurable characterization theory to promote their quantum simulations.
 
In this Letter, we propose a concept of polarized topological charges in momentum space to characterize the chiral-symmetric HOTPs, with which an experimental scheme to realize and detect these HOTPs in $^{87}$Rb cold atomic system is provided. Our characterization scheme has theoretical simplicity and observational intuitiveness based on the following nontrivial discoveries. Firstly, the second-order (third-order) topological phases are determined by a quarter (negative eighth) of the total polarized topological charges. Secondly, the polarized topological charges are identified by measuring the pseudospin structures of the systems. Finally, this theory exactly distinguishes two types of topological phase transitions induced by closing the band energy gap of bulk states or edge states, completely covers the different topological classifications of chiral-symmetric HOTPs, and has broad applications. These results shall advance research of the HOTPs in both theory and experiment. 

{\it\color{blue}Generic theory in 2D systems.}---Our starting point is a family of 2D systems that host chiral-symmetric second-order topological phases. The minimal momentum-space Hamiltonian to satisfy this requirement reads
\begin{equation}
{\cal H}(\mathbf{k})=\sum_{i=1}^4h_i(\mathbf{k})\Gamma_i.
\label{eq:ChiralHamiltonian}
\end{equation}
The Gamma matrices obeying $\{\Gamma_i,\Gamma_j\}=2\delta_{ij}$ are of dimensionality $4$ (classes AIII and BDI) or $8$ (class CII)~\cite{Lei2022,luo2022higher}, which guarantees that there is a chiral operator $\Gamma_5$ to satisfy $\Gamma_5{\cal H}(\mathbf{k})\Gamma_5= -{\cal H}(\mathbf{k})$. We hereby arrange the Gamma matrices in an order satisfying the trace property $\text{Tr}(\Gamma_5\Gamma_1\Gamma_2\Gamma_3\Gamma_4)=-4$~\cite{morimoto2013topological,chiu2013classification}. Hence the systems have chiral symmetry and possess $\mathbb{Z}$-classified second-order topological phases~\cite{benalcazar2017quantized,langbehn2017reflection,wu2020boundary,ezawa2020edge,li2020topological,benalcazar2022chiral,Lei2022,luo2022higher}.

\begin{figure}[t!]
\centering
\includegraphics[width=1.0\columnwidth]{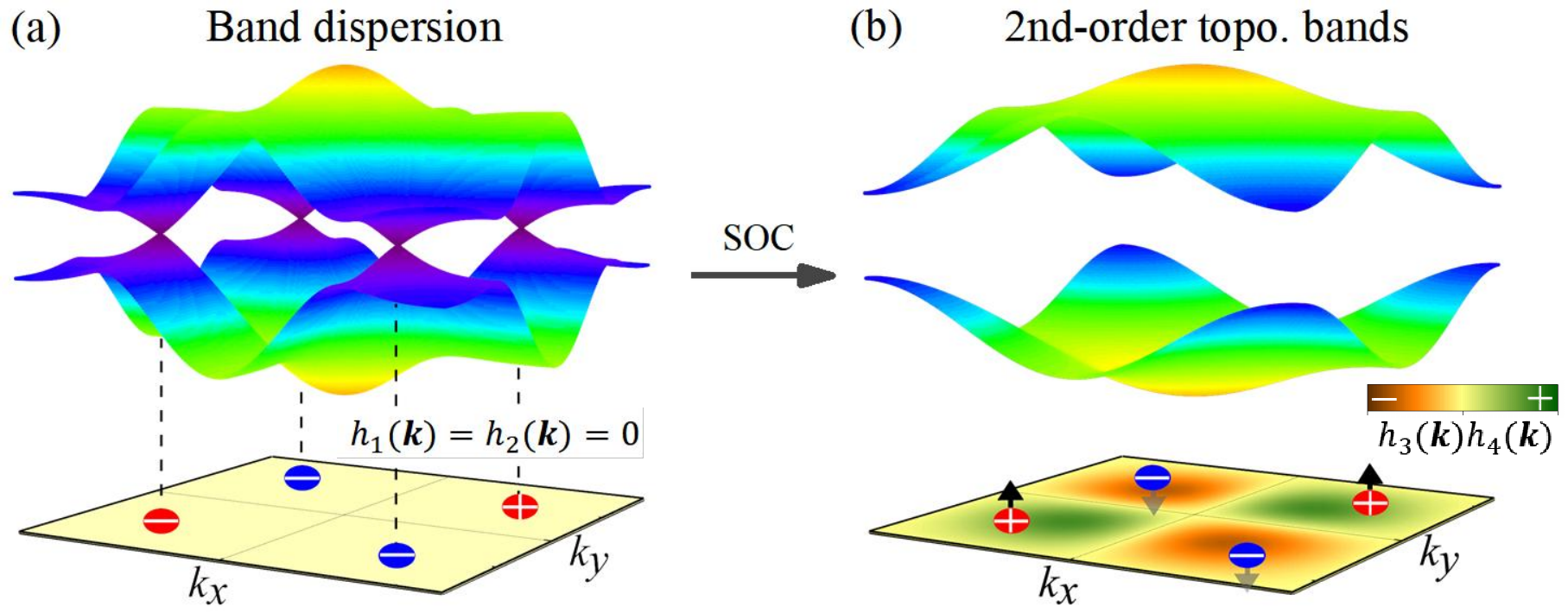}
\caption{(a) 2D band dispersion with $E_b(\mathbf{k})=\pm(h_1^2+h_2^2)^{1/2}$ gives four nodal points with positive or negative unit-value topological charges. (b) The SO coupling opens band gap and two energy bands become $E_g(\mathbf{k})=\pm(h_1^2+h_2^2+h_3^2+h_4^2)^{1/2}$, which causes the topological charges to be polarized. The charge polarizations ${\cal P}_n=+1$ and $-1$ are marked by arrows parallel and antiparallel to the direction ${\bf k}_x\times{\bf k}_y$, respectively.}
\label{fig:1}
\end{figure}

\begin{figure*}[ht]
\centering
\includegraphics[width=2.0\columnwidth]{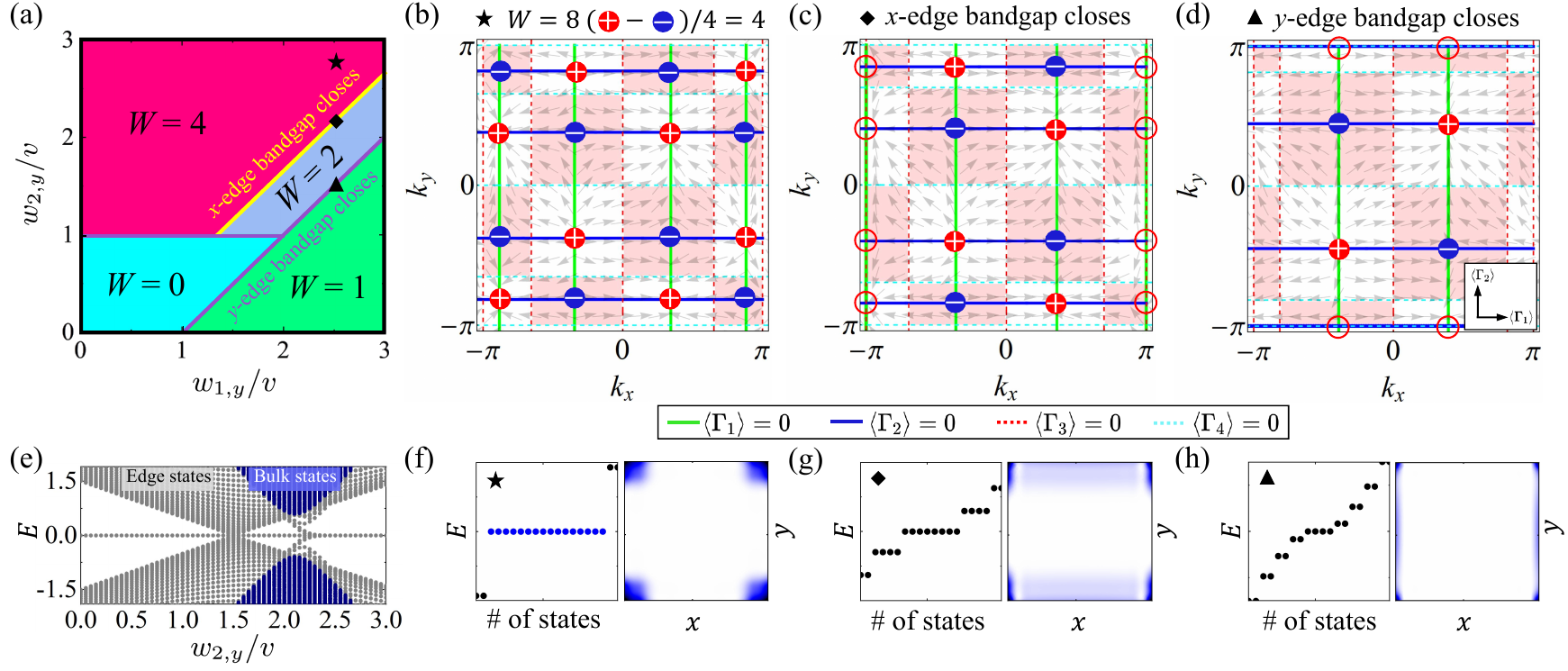}
\caption{(a) Phase diagram indicated by $W$. Pseudospin field $\boldsymbol{\Theta}(\mathbf{k})$ with the marked topological charges $\mathcal{C}_n$ and polarization $\mathcal{P}_n=+1$ in the white regions and $-1$ in the pink ones for (b) $W=4$ when $w_{2,y}=2.7v$ (star), and for (c) $x$- and (d) $y$-edge bandgap closing points when $w_{2,y}=2.167v$ (rhombus) and $1.5v$ (triangle), respectively. Open red circles mark where the topological charges are annihilated or created. (e) Energy spectrum under the open-boundary condition in the different $w_{2,y}$ when $w_{1,y}=2.5v$. The topological phase transitions occur at closing points of the edge-state band gap. (f)-(h) Zero-energy states and their real-space distributions for (b)-(d). We use $w_{1(2),x}=3w_{1(2),y}$ and $w_{1,y}=2.5v$.}
\label{fig:2}
\end{figure*}

Inspired by the description of first-order topology~\cite{sticlet2012geometrical,zhang2018dynamical,zhang2019dynamical,jia2020charge}, we resort to the topological charge defined in momentum space to characterize the second-order topological phases. Specifically, we separate $\mathbf{h}({\bf k})$ of Eq.~\eqref{eq:ChiralHamiltonian} into two parts, i.e., $\mathbf{h}({\bf k})=(\mathbf{h}_\text{m},\mathbf{h}_\text{so})$, where $\mathbf{h}_\text{m}=(h_1,h_2)$ describes the band dispersion and $\mathbf{h}_\text{so}=(h_3,h_4)$ plays a similar role as the pseudospin-orbit (SO) coupling. In the absence of $\mathbf{h}_\text{so}$, there are two gapless bands [see Fig.~{\ref{fig:1}}(a)]. The $n$th nodal point $\boldsymbol{\varrho}_n$ is determined by $\mathbf{h}_\text{m}(\boldsymbol{\varrho}_n)=0$. A topological charge at $\boldsymbol{\varrho}_n$ is then defined as
\begin{align}
{\cal C}_n=\text{sgn}\left[\mathsf{J}_{\mathbf{h}_\text{m}}(\boldsymbol{\varrho}_n)\right],~~\mathsf{J}_{\mathbf{h}_\text{m}}(\mathbf{k})=\text{det}\left[\partial{h_{\text{m},i}(\mathbf{k})}/\partial{k_j}\right],
\label{eq:TopologicalCharge}
\end{align}
which describes the winding of $\mathbf{h}_\text{m}$ around $\boldsymbol{\varrho}_n$. The presence of $\mathbf{h}_\text{so}$ opens an energy gap and these nodal points are gapped out [see Fig.~{\ref{fig:1}}(b)]. We then endow the topological charge of $\boldsymbol{\varrho}_n$ with a polarization  
\begin{align}
{\cal P}_n=\text{sgn}\left[\mathsf{P}(\boldsymbol{\varrho}_n)\right],~~\mathsf{P}(\mathbf{k})=\int\mathsf{J}_{\mathbf{h}_\text{so}}(\mathbf{k})d{\mathbf{k}}.
\label{eq:ChargePolzrization}
\end{align}
The topological invariant of this system is defined as a quarter of total polarized topological charges in the BZ
\begin{align}
W=\frac{1}{4}\sum_n{\cal C}_n{\cal P}_n,
\label{eq:TopologicalCharacterization}
\end{align}
which provides a momentum-space characterization to the second-order topological phases. In the topologically nontrivial regime, each nodal point has a nonzero ${\cal C}_n$ and ${\cal P}_n$, which contributes a nontrivial $W$ [see Fig.~\ref{fig:2}(a)]. With changing system parameters, the bulk-state or edge-state band gap closes. The new (original) topological charges are created (annihilated) in the zero polarization regions. This causes an abrupt change of $W$ and signifies the emergence of topological phase transitions, as we show it in Figs.~\ref{fig:2}(c) and \ref{fig:2}(d). It is noted that this characterization works not only for the generic separable systems but also for certain inseparable systems~\cite{SuppInfo}. 
 
The topological invariant $W$ is actually an extension of the 1D winding number. To demonstrate this result, we first consider a Su–Schrieffer–Heeger (SSH) chain along $x$ direction. The corresponding momentum-space Hamiltonian reads ${\cal H}(k_x)=h_1(k_x)\tau_x+h_3(k_x)\tau_y$ and has chiral symmetry under the chiral operator $\tau_z$, where $\tau_{x,y,z}$ are Pauli matrices. One can define a 1D winding number $W_x=(1/2\pi \mathtt{i})\int_{\text{BZ}}\text{Tr}\left[q(k_x)^\dagger\partial_{k_x}q(k_x)\right]dk_x$, with $q(k_x)=h_1(k_x)+\mathtt{i}h_3(k_x)$, to characterize its first-order topology~\cite{chiu2016classification}. Remarkably, we find that this winding number equals exactly to $W_x=(-1/2)\sum_{l_x=1}^{N_x}{\cal C}^x_{l_x}{\cal P}^x_{l_x}$~\cite{SuppInfo}, where ${\cal C}^x_{l_x}=\text{sgn}[\partial h_1(\varrho_{l_x})/\partial{k_x}]$ are the topological charges, ${\cal P}^x_{l_x}=\text{sgn}\left[h_3(\varrho_{l_x})\right]$ are the charge polarizations, and $N_x$ is the number of the nodal points. When a $y$-directional SSH chain described by ${\cal H}(k_y)=h_2(k_y)\sigma_x+h_4(k_y)\sigma_y$, which possesses a similar winding number $W_y$ to $W_x$, is further stacked to ${\cal H}(k_x)$, we obtain a special 2D system described by ${\cal H}(k_x,k_y)={\cal H}(k_x)\otimes 1+\tau_z\otimes{\cal H}(k_y)$, whose variables $k_x$ and $k_y$ are separable. Being equivalent to the Benalcazar-Bernevig-Hughes (BBH) model~\cite{benalcazar2017quantized,benalcazar2017electric}, it hosts the chiral-symmetric second-order topological phases. According to Ref.~\cite{okugawa2019second}, the second-order topology of this $2$D system is characterized by $W=W_xW_y$, which is just Eq.~\eqref{eq:TopologicalCharacterization} by redefining the topological charges ${\cal C}_n={\cal C}^x_{l_x}{\cal C}^y_{l_y}$ and their polarizations ${\cal P}_n={\cal P}^{x}_{l_x}{\cal P}^y_{l_y}$ for the nodal points $\boldsymbol{\varrho}_n=(\varrho_{l_x},\varrho_{l_y})$, where $l_{x/y}=1,2,\cdots,N_{x/y}$ and $n=1,2,\cdots, N_xN_y$. Second, we can convert the above system to a generic separable system by an orthogonal transformation to the Gamma matrices without changing its topology~\cite{SuppInfo,trifunovic2021higher}. It makes our characterization also applicable in the generic separable systems, whose topological charges and charge polarizations are generalized to Eq.~\eqref{eq:TopologicalCharge} and Eq.~\eqref{eq:ChargePolzrization}, respectively. Finally, our scheme can also be applied in the 2D inseparable systems ${\cal H}({\bf k})$ possessing mirror-rotation symmetry. Their second-order topology is described by ${\cal H}(k)$ on its high-symmetric lines $k_x=k_y=k$~\cite{liu2019second}, which is obviously separable. It implies that $W$ is applicable in these inseparable systems~\cite{SuppInfo}. This result has be verified by the model of Eq.~\eqref{eq:EBHZ}. Therefore, we provide an elegant characterization for the chiral-symmetric second-order topological phases of 2D systems. 

{\it\color{blue}Generalizations in $d$D systems and Momentum-space measurements.}---The above theory can be generalized to characterize the $d$D $d$th-order chiral-symmetric topological phases described by ${\cal H}(\mathbf{k})=\sum^{2d}_{i=1} h_i(\mathbf{k})\Gamma_i$, which are constructed by the anticommuting Gamma matrices with the dimensionality $2^d$ for the classes AIII and BDI or $2^{d+1}$ for the class CII. Thus the chiral operator $\Gamma_{2d+1}$ is guaranteed and we have the trace property of $\text{Tr}(\Gamma_{2d+1}\prod^{2d}_{i=1}\Gamma_i)=(-2\mathtt{i})^{d}$. In a similar manner to the 2D case, we separate $\mathbf{h}({\bf k})=(\mathbf{h}_\text{m},\mathbf{h}_\text{so})$, with $\mathbf{h}_\text{m}=(h_1,h_2,\cdots,h_d)$ and $\mathbf{h}_\text{so}=(h_{d+1},h_{d+2},\cdots h_{2d})$. The $d$th-order topology is characterized by
\begin{equation}
W=\frac{1}{(-2)^d}\sum_n{\cal C}_n{\cal P}_n,
\end{equation}
where ${\cal C}_n=\text{sgn}\{\text{det}\left[\partial{h_{\text{m},i}(\boldsymbol{\varrho}_n)}/\partial{k_j}\right]\}$ are the topological charges located at the nodal points $\boldsymbol{\varrho}_n$ and determined by $\mathbf{h}_\text{m}(\boldsymbol{\varrho}_n)=0$~\cite{SuppInfo}. The charge polarizations reads ${\cal P}_n=\int\mathsf{J}_{\mathbf{h}_\text{so}}(\mathbf{k})d{\mathbf{k}}$. Accordingly, the different types of $d$th-order topological phase transitions are captured by the emergence or merger of topological charges in the zero polarization regions. 

This $W$ is observable by measuring the pseudospin expectation $\langle\Gamma_i(\mathbf{k})\rangle=\bra{u(\mathbf{k})}\Gamma_i\ket{u(\mathbf{k})}$, where $\ket{u(\mathbf{k})}$ is the ground state of $\mathcal{H}({\bf k})$. We further define a pseudospin field $\boldsymbol{\Theta}(\mathbf{k})$, whose component is $\Theta_i(\mathbf{k})=-{\langle\Gamma_i(\mathbf{k})\rangle}/{{{\cal N}_\mathbf{k}}}$, with $i=1,2,\cdots ,d$ and ${\cal N}_\mathbf{k}$ being a normalization factor. Using the anticommutation relation of $\boldsymbol{\Gamma}$, we obtain $\langle\Gamma_i(\mathbf{k})\rangle=-h_i/(\sum_{i=1}^{2d}h_i^2)^{1/2}$. The SO couplings generally have linear dispersion, i.e., $h_{d+i}(\mathbf{k})=h_{d+i}(k_{i})$, with $i=1,2,\cdots,d$. It readily leads to ${\cal P}_n=\text{sgn}\left[h_{d+1}(\boldsymbol{\varrho}_n)\cdots h_{2d}(\boldsymbol{\varrho}_n)\right]=\text{sgn}\left[(-1)^d\langle\Gamma_{d+1}(\boldsymbol{\varrho}_n)\rangle\cdots\langle\Gamma_{2d}(\boldsymbol{\varrho}_n)\rangle\right]$ and $\boldsymbol{\Theta}(\boldsymbol{\varrho}_n)= \mathbf{h}_\text{m}(\boldsymbol{\varrho}_n)$. The topological charge ${\cal C}_n$ is recast into the winding of $\boldsymbol{\Theta}(\mathbf{k})$ around the nodal points $\boldsymbol{\varrho}_n$, which reads ${\cal C}_n=\text{sgn}\{\text{det}[(\partial\Theta_i(\boldsymbol{\varrho}_n)/\partial k_j)]\}$. Since the pseudospin structures of $\langle\Gamma_i(\mathbf{k})\rangle$ is measurable in quantum simulation experiments~\cite{ji2020quantum,yu2020high,niu2020simulation}, our scheme provides an insightful picture to detect the chiral-symmetric HOTPs in the momentum space.

{\it\color{blue}Applications in typical models.}---Our characterization theory can be applied to the different classes of topological systems. We first consider a separable system whose topological phase transition occurs at the closing points of the edge-state band gap instead of the bulk one. It is an extended BBH model~\cite{benalcazar2022chiral} with
\begin{equation}
\begin{split}
&h_{1(2)}=v+ w_{1,x(y)}\cos k_{x(y)}+ w_{2,x(y)}\cos 2k_{x(y)},\\
&h_{3(4)}=w_{1,x(y)}\sin k_{x(y)}+w_{2,x(y)}\sin 2k_{x(y)}.\\
\end{split}\label{sepse}
\end{equation}
The $\boldsymbol{\Gamma}$ matrices are $\Gamma_1=-\tau_x\sigma_z$, $\Gamma_2=\tau_x\sigma_x$, $\Gamma_3=-\tau_y$, and $\Gamma_4=-\tau_x\sigma_y$. It has chiral symmetry under $\Gamma_5=\tau_z$, mirror symmetry with $M_{x}=\tau_x$ and $M_{y}=\tau_y\sigma_y$, and inversion symmetry with $I=\tau_z\sigma_y$. The system belongs to the topological class AIII. We show in Fig.~\ref{fig:2}(a) the phase diagram described by $W$ via studying the expectation values $\langle\Gamma_i(\mathbf{k})\rangle$. When $w_{2,y}=2.7v$, there are sixteen nodal points determined by $\boldsymbol{\Theta}(\mathbf{k})=0$ [see Fig.~\ref{fig:2}(b)]. Their topological charges $\mathcal{C}_n$ are identified by calculating the winding of $\boldsymbol{\Theta}(\mathbf{k})$ around $\boldsymbol{\varrho}_n$. The signs of $\langle\Gamma_{3,4}(\boldsymbol{\varrho}_n)\rangle$ determine the polarizations, where ${\cal P}_n=+1$ and $-1$ are marked by the white and pink, respectively. We immediately have $W=4$, which signifies that the system hosts sixteen degenerate zero-energy states localized at its corners [see Fig.~\ref{fig:2}(f)]. Figure~\ref{fig:2}(c) shows $\langle\Gamma_i({\bf k})\rangle$ and the polarizations at the phase boundary between $W=4$ and $2$ when $w_{2,y}=2.167v$. Four pairs of nodal points satisfying $\langle\Gamma_{1,2}(\boldsymbol{\varrho}_n)\rangle=0$ merge to the line of $k_x=-\pi$ of $\langle\Gamma_3({\bf k})\rangle=0$, which, keeping the band gap of the bulk states open due to $\langle\Gamma_4({\bf k})\rangle\neq0$, closes the band gap of $x$-edge states [see Figs.~\ref{fig:2}(e) and \ref{fig:2}(g)]. It makes that the polarization of these nodal points vanish and a topological phase transition is triggered. A similar case of the phase boundary with a gap closing of the $y$-edge states is given in Figs.~\ref{fig:2}(d) and \ref{fig:2}(h).

\begin{figure}[t]
\centering
\includegraphics[width=1.0\columnwidth]{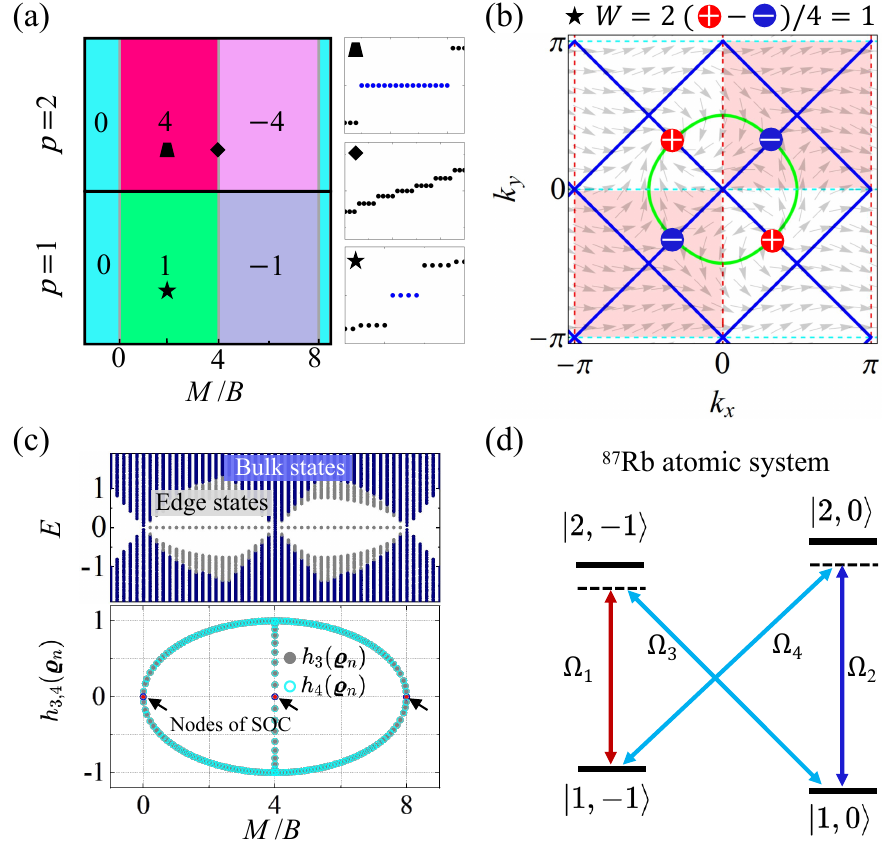}
\caption{(a) Phase diagrams indicated by $W$ and zero-energy states when $(p,M/B)=(2,2.0)$ (trapezoid), $(2,4.0)$ (rhombus), and $(1,2.0)$ (star). (b) Pseudospin structures of $\boldsymbol{\Theta}(\mathbf{k})$ with the marked $\mathcal{C}_n$ and polarization for $W=1$ when $(p,M/B)=(1,2.0)$. (c) Energy spectrum under the open-boundary conditions (upper). The numerical results of $h_{3,4}(\boldsymbol{\varrho}_n)$ (lower), where $h_{3,4}(\boldsymbol{\varrho}_n)=0$ (red solid points and blue open cycles) emerging at $M=0$, $4B$, and $8B$ gives the topological phase transitions driven by bulk-state band gap closing. (d) Simulation of the Hamiltonian of class BDI by a four-level atomic system.}
\label{fig:3}
\end{figure}

We further consider an inseparable system whose topological phase transition occurs at the closing point of the bulk-state band gap. Its $h$-components are
\begin{eqnarray}
h_1&=&M-2B[2-\cos (pk_x)-\cos (pk_y)],~h_3=\sin( pk_x),\nonumber\\
h_2&=&\cos (2pk_x) -\cos (2pk_y),~h_4=-\sin (pk_y),
\label{eq:EBHZ}
\end{eqnarray}
with an integer $p$. The $\boldsymbol{\Gamma}$ matrices are $\Gamma_1=\sigma_z$, $\Gamma_2=\tau_x\sigma_x$, $\Gamma_3=\tau_z\sigma_x$, and $\Gamma_4=\sigma_y$. The system has chiral symmetry under $\Gamma_5=\tau_y\sigma_x$, time-reversal symmetry with ${\cal T}=\tau_x {\cal K}$, and particle-hole symmetry with $P=\tau_z\sigma_x{\cal K}$, where $\cal K$ is complex conjugate operator. Therefore, it belongs to the class BDI. In addition to the symmetries of $M_x=\tau_y\sigma_z$, $M_y=\tau_x$, $I=\tau_z\sigma_z$, the system also has $C_4$-rotation symmetry with $C_4=\text{diag}(\mathtt{i},1,-1,-\mathtt{i})$ and mirror-rotation symmetry with $M_{xy}=C_4M_y$. The phase diagram characterized by $W$ is shown in Fig.~\ref{fig:3}(a). The structures of $\boldsymbol{\Theta}(\mathbf{k})$ when $(p,M/B)=(1,2.0)$ figure out four nodal points, as shown in Fig.~\ref{fig:3}(b). Via calculating the winding of $\boldsymbol{\Theta}(\mathbf{k})$ around $\boldsymbol{\varrho}_n$, we have the topological charge $\mathcal{C}_n$ of $\boldsymbol{\varrho}_n$. Their polarization are achieved by examining the signs of $\langle\Gamma_{3,4}(\boldsymbol{\varrho}_n)\rangle$. We then readily obtain $W=1$, which signifies that four corner states are present.  The energy spectrum under the open-boundary condition in Fig.~\ref{fig:3}(c) reveals that the phase transition occurs at the closing points of the band gap of the bulk states. Such a phase transition is caused by the vanishing of the charge polarizations $\mathcal{P}_n$ driven by $h_{3,4}(\boldsymbol{\varrho}_n)=0$. 

Besides, keeping $h$-components as Eq.~(\ref{eq:EBHZ}) but changing the $\boldsymbol{\Gamma}$ matrices as $\Gamma_1=\tau_x$, $\Gamma_2=\rho_z\tau_z\sigma_x$, $\Gamma_3=\rho_z\tau_y$, $\Gamma_4=\rho_z\tau_z\sigma_y$, and $\Gamma_5=\rho_z\tau_z\sigma_z$, we shall obtain a second-order topological phase belonging to class CII, where $2W$ gives the second-order topology of the corner states and we can identify $W$ like the BDI case. We also provide the simulation results of the 3D BBH model in Supplementary Materials~\cite{SuppInfo}, which further demonstrates the broad applicability of this theory.

\begin{figure}[t]
\centering
\includegraphics[width=1.0\columnwidth]{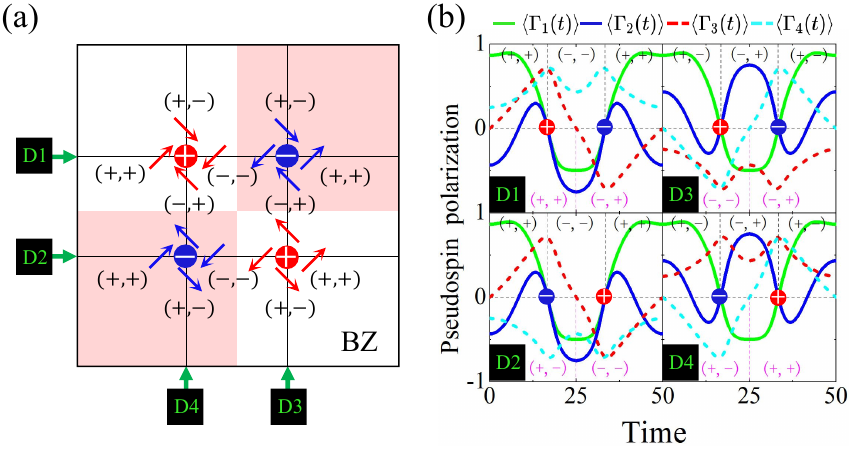}
\caption{(a) Detecting polarized topological charges by sweeping $k_{x(y)}$ at four lines $\text{D}_{1,2,3,4}$ of BZ. (b) Dynamical pseudospin polarizations $\langle\Gamma_i(t)\rangle$ of $p=1$, where $\langle\Gamma_{1,2}(t)\rangle=0$ give the locations of the nodal points. Signs of $\langle\Gamma_{1,2}(t)\rangle$ and $\langle\Gamma_{3,4}(t)\rangle$ near at the nodal points are marked by black and pink, determining their charges and polarizations, respectively.}
\label{fig:4}
\end{figure}

{\it\color{blue}Experimental scheme for realizing HOTPs.}---Based on the realization of Bernevig-Hughes-Zhang (BHZ) model in the recent experiment~\cite{lv2021measurement}, we provide a scheme to realize the 2D chiral-symmetric second-order topological phases in $^{87}$Rb cold atomic system. We take $p=1$ as an example to demonstrate our scheme. By employing four atomic hyperfine levels $\ket{a}=\ket{2,-1}$, $\ket{b}=\ket{1,-1}$, $\ket{c}=\ket{2,0}$, and $\ket{d}=\ket{1,0}$, we apply four microwaves to couple $\{\ket{a},\ket{b}\}$, $\{\ket{c},\ket{d}\}$, $\{\ket{a},\ket{d}\}$, and $\{\ket{c},\ket{b}\}$ with Rabi frequencies $\Omega_1=-\Omega_2=(h_3^2+h_4^2)^{1/2}$ and $\Omega_3=\Omega_4=h_2$ and phases $\varphi_1=-\varphi_2=\arctan (h_4/h_3)$ and $\varphi_3=\varphi_4=0$ [see Fig.~\ref{fig:3}(c)]. Then, Eq.~(\ref{eq:EBHZ}) is obtained on the bare-state basis $\{\ket{a},\ket{b},\ket{c},\ket{d}\}$~\cite{SuppInfo}. It is found that the $\mathbf{k}$-dependent $h_2$ term opens band gap of the helical states of the BHZ model and induces the corner states, rendering a second-order topological phases.

The topological charges and their charge polarizations can be detected by the following dynamical way \citep{lv2021measurement}. Choosing one fixed $k_{y(x)}$ and sweeping $k_{x(y)}$ at a rate $v_{k_{x(y)}}=2\pi/T$, i.e., $k_{x(y)}(t)=v_{k_{x(y)}}t-\pi$ with $t\in[0,T]$, we perform the measurements along four different momentum lines $\text{D}_{1,2,3,4}$ of BZ, which covers all the topological charges [see Fig.~\ref{fig:4}(a)]. It is seen that the nodal points are captured by the dynamical expectation values $\langle\Gamma_{1}(t)\rangle=\langle\Gamma_{2}(t)\rangle=0$ [see Fig.~\ref{fig:4}(b)]. The topological charges and charge polarizations are determined by the signs of $\langle\Gamma_{1,2}(t)\rangle$ and $\langle\Gamma_{3,4}(t)\rangle$ near at these nodal points, respectively [see Figs.~\ref{fig:4}(a) and \ref{fig:4}(b)]. We can finally identify $W=1$ in this second-order topological system. These results provide possibility to observe the chiral-symmetric HOTPs in quantum simulations.

{\it\color{blue}Discussion and Conclusion.}---Although we have studied the chiral-symmetric HOTPs in static and Hermitian systems, our topological characterization theory may have broad implications in Floquet HOTPs~\cite{hu2020dynamical} or non-Hermitian HOTPs~\cite{liu2019second}. When adding the periodic driving or non-Hermitian terms in the chiral-symmetric HOTPs, these $\mathbb{Z}_2$ topological invariants in Refs.~\cite{hu2020dynamical} and \cite{liu2019second} can not characterize these topological states. Generalizing the real-space topological invariants~\cite{benalcazar2022chiral,lin2024probing,li2024exact} in these systems also becomes exceptionally difficult. Nevertheless, the polarized topological charges may provide an possible way to solve these interesting questions. Indeed, the topological charge has been employed to describe the Floquet topological states~\cite{zhang2020unified,zhang2022unconventional,wang2024characterizing} and promoted the realization and detection of these Floquet topological states in the ultracold cold atoms~\cite{zhang2023tuning}. Besides, our topological characterization theory can be extended the more general models where the dimensionality of Gamma matrices are $2^{d+j}$, with the positive integer $j$. When keeping the expressions of $h$-components and the topological classifications of bulk Hamiltonians, the number of corner zero modes shall increase, which are characterized by $(j+1)\mathbb{Z}$-classified topological invariants. Our theory still exactly describes these topological states via the topological index $(j+1)W$, with $W=1/(-2)^d\sum_n\mathcal{C}_n\mathcal{P}_n$. In summary, we have developed the concept of polarized topological charge, with which an experimentally measurable momentum-space topological characterization for the chiral-symmetric HOTPs is proposed. This characterization theory further inspires a feasible experimental realization and detection to the HOTPs in the $^{87}$Rb cold atomic system. Our work promotes the theoretical study of HOTPs and shall stimulate their experimental realization and detection in realistic quantum simulation platforms.

\emph{Acknowledgements.}  We thank Yucheng Wang and Long Zhang for the helpful discussions. This work is supported by the National Natural Science Foundation (Grants No. 12275109, No. 12247101, and No. 12404318).

\bibliographystyle{apsrev4-1}

\pagebreak
\clearpage
\onecolumngrid
\flushbottom
\begin{center}
\textbf{\large Supplementary Material for ``Unveiling higher-order topology via polarized topological charges"}
\end{center}
\setcounter{equation}{0}
\setcounter{figure}{0}
\setcounter{table}{0}
\makeatletter
\renewcommand{\theequation}{S\arabic{equation}}
\renewcommand{\thefigure}{S\arabic{figure}}
\renewcommand{\bibnumfmt}[1]{[S#1]}
\renewcommand{\citenumfont}[1]{S#1}

In this Supplementary Material, we provide the details of polarized topological charge in first-order and higher-order topological systems in Secs. \ref{fstod} and \ref{hodtp}, respectively. We also show the numerical results for second-order and third-order topological phases with chiral symmetry in Sec. \ref{nul}. We further give the higher-order topological phase transitions determined by the polarized topological charges in Sec. \ref{hotpt}. We finally provide the experimental scheme in $^{87}$Rb atomic system to realize the chiral-symmetric topological phase with corner states in Sec. \ref{exp}.

\section{Polarized topological charges in first-order topological systems}\label{fstod}
We first propose the concept of polarized topological charge in the first-order topological systems. Our starting point is a one-dimensional (1D) chiral-symmetric topological phases in the $\mathbb{Z}$ topological classification, such as a Su–Schrieffer–Heeger (SSH) model along $x$ direction. The corresponding momentum-space Hamiltonian is written as 
\begin{equation}
{\cal H}(k_x)=h_1(k_x)\tau_x+h_3(k_x)\tau_y,
\label{x_SSH_model}
\end{equation}where $\tau_{x,y,z}$ are Pauli matrices.
This system has chiral symmetry $\tau_z{\cal H}(k_x)\tau_z={\cal H}(-k_x)$ under the chiral operator $\tau_z$. Thus, its first-order topology is characterized by the winding number 
\begin{equation}
W_x=\frac{1}{2\pi i}\int_{\text{BZ}}\text{Tr}\left[q(k_x)^\dagger\partial_{k_x}q(k_x)\right],
\label{windingnumber}
\end{equation}
where $q$ is a unitary matrix and defines a map from the 1D Brillouin zone (BZ) to the space of unitary matrices $U(n)$. This gives an integer topological classification by the first homotopy group, i.e., $\pi_1[U(n)]=\mathbb{Z}$. Hence $W_x$ is an integer topological invariant to describe the first-order topological phase with chiral symmetry. Remarkably, $W_x$ can also be determined by the characterization scheme with polarized topological charges. We use the convention that $h_1(k_x)$ denotes the band dispersion, while $h_3(k_x)$ denotes the pseudospin-orbit (SO) coupling. $N_x$ nodal points are found from the band dispersion $h_1(\varrho_l)=0$. We can define a topological charge to each $\varrho_l$ as 
\begin{equation}
{\cal C}^x_l=\text{sgn}\Big[\frac{\partial h_1(\varrho_l)}{\partial{k_x}}\Big].\label{xcharges}
\end{equation}
The presence of the pseudospin-orbit coupling term opens the band gap and make the topological charge $\mathcal{C}_l$ polarized in an amount
\begin{equation}
{\cal P}^x_l=\text{sgn}\left[h_3(\varrho_l)\right]. 
\label{xpolarization}
\end{equation}
Then, one can analytically prove that Eq. \eqref{windingnumber} is equivelent to 
\begin{equation}
W_x=-\frac{1}{2}\sum_{l=1}^{N_x}{\cal C}^x_{l}{\cal P}^x_{l},\label{fstodtp}
\end{equation}
where ${\cal C}^x_{l}{\cal P}^x_l$ is called a polarized topological charge. It should be noted that we hereby use the properties of all the polarized topological charges in the BZ to characterize $W_x$. Physically, each ${\cal C}^x_l$ is defined on a nodal point of the band dispersion, and then the nonzero SO coupling opens energy gap at this nodal point, which makes ${\cal C}^x_l$ obtain a nonzero polarization ${\cal P}^x_l$. This tells us that all the polarized topological charges of the BZ are nonzero for a gapped topological phase, while the emergence of one zero-value polarized topological charge can drive the topological phase transition, where the energy band gap is closing at this nodal point. 

\section{Polarized topological charges in higher-order topological systems}\label{hodtp}

\subsection{Special $2$D separable systems}

Next we shall generalize the concept of polarized topological charge to higher-order topological systems. We first consider the special 2D lattice systems hosting second-order topological phases within the $\mathbb{Z}$ classification, such as Benalcazar-Bernevig-Hughes (BBH) model~\cite{benalcazar2017quantized-s} or 2D SSH model~\cite{liu2017novel-s}, which can be obtained by stacking $y$-directional SSH along $x$ direction. Its momentum-space Hamiltonian is written as 
\begin{equation}
{\cal H}(k_x,k_y)=h_1(k_x)\Gamma_1+h_2(k_y)\Gamma_2+h_3(k_x)\Gamma_3+h_4(k_y)\Gamma_4.
\label{special_sep}
\end{equation}
This Hamiltonian can be translated to a separable form ${\cal H}(k_x,k_y)={\cal H}(k_x)\otimes 1+\tau_z\otimes{\cal H}(k_y)$ for a BBH model or ${\cal H}(k_x,k_y)={\cal H}(k_x)\otimes \sigma_y+1\otimes{\cal H}(k_y)$ for a 2D SSH model. Due to separability of $k_x$ and $k_y$ in Eq. \eqref{special_sep}, its second-order topology can be characterized by $W=W_xW_y$~\cite{okugawa2019second-s}. Then, from Eq. \eqref{fstodtp} and the similar form of $W_y$, we have
\begin{equation}
W=\frac{1}{4}\sum_{l=1}^{N_x}{\cal C}^{x}_l{\cal P}^{x}_l\sum_{m=1}^{N_y}{\cal C}^y_m{\cal P}^y_m. 
\label{topocharacterization}
\end{equation}
Redefining the topological charges for the $N_xN_y$ nodal points $(\varrho_l,\varrho_m)$ as
\begin{align}
{\cal C}_n={\cal C}^x_l{\cal C}^y_m=\text{sgn}\left[\begin{vmatrix}
\partial_{k_x}h_1(\varrho_l) & 0\\
0 & \partial_{k_y}h_2(\varrho_m)
\end{vmatrix}\right],
\label{seq:chargeseparability}
\end{align}
and the corresponding polarizations as ${\cal P}_n={\cal P}^{x}_l{\cal P}^y_m=\text{sgn}[h_3(\varrho_l)h_4(\varrho_m)]$, Eq. \eqref{topocharacterization} is recast into
\begin{eqnarray}
W&=&\frac{1}{4}({\cal C}^{x}_{1}{\cal C}^{y}_{1}{\cal P}^{x}_1{\cal P}^{y}_1+{\cal C}^{x}_{1}{\cal C}^{y}_{2}{\cal P}^{x}_1{\cal P}^{y}_2\cdots+{\cal C}^{x}_{N_x}{\cal C}^{y}_{N_y}{\cal P}^{x}_{N_x}{\cal P}^{y}_{N_y})\nonumber\\
&=&\frac{1}{4}({\cal C}_{1}{\cal P}_1+{\cal C}_{2}{\cal P}_2\cdots+{\cal C}_{N_x\times N_y}{\cal P}_{N_x\times N_y})\nonumber\\
&=&\frac{1}{4}\sum^{N_x\times N_y}_{n=1}{\cal C}_{n}{\cal P}_{n}.
\label{narray}
\end{eqnarray}
Thus, the special $2$D separable system is well described by the polarized topological charges. 

\subsection{Generic $2$D separable systems}

Next we generalize the polarized topological charges to the generic 2D separable systems, in which we use a key idea that the generic separable Hamiltonian can be deformed to the special separable Hamiltonian without changing topology~\cite{trifunovic2021higher-s}. To demonstrate this point, we start from the following Hamiltonian 
\begin{equation}
{\cal H}(k_x,k_y)=\frac{g_1(k_x)-g_2(k_y)}{\sqrt{2}}\Gamma_1+\frac{g_1(k_x)+g_2(k_y)}{\sqrt{2}}\Gamma_2+\frac{g_3(k_x)-g_4(k_y)}{\sqrt{2}}\Gamma_3+\frac{g_3(k_x)+g_4(k_y)}{\sqrt{2}}\Gamma_4,
\label{sep}
\end{equation}
which is obviously not a special separable Hamiltonian as Eq. \eqref{special_sep}. By taking $\tilde{\Gamma}_1=\frac{\Gamma_1+\Gamma_2}{\sqrt{2}}$, $\tilde{\Gamma}_2=\frac{\Gamma_2-\Gamma_1}{\sqrt{2}}$, $\tilde{\Gamma}_3=\frac{\Gamma_3+\Gamma_4}{\sqrt{2}}$, and $\tilde{\Gamma}_4=\frac{\Gamma_4-\Gamma_3}{\sqrt{2}}$, Eq. \eqref{sep} is rewritten as 
\begin{equation}
{\cal H}(k_x,k_y)=g_1(k_x)\tilde{\Gamma}_1+g_2(k_y)\tilde{\Gamma}_2+g_3(k_x)\tilde{\Gamma}_3+g_4(k_y)\tilde{\Gamma}_4.
\label{insep}
\end{equation}
Here we have $\{\tilde{\Gamma}_i,\tilde{\Gamma}_j\}=\delta_{ij}$, which obey the anticommutation relation of the Clifford algebra. Equation~\eqref{insep} becomes a similar separable form to Eq.~\eqref{special_sep}. Hence, the topological charges are written as 
\begin{equation}
{\cal C}_n=\text{sgn}[\mathsf{J}_{\mathbf{h}_\text{m}}(\boldsymbol{\varrho}_n)]=\text{sgn}[\begin{vmatrix}
\partial_{k_x}g_1(k_x) & 0\\
0 & \partial_{k_y}g_2(k_y)
\end{vmatrix}]
\end{equation}
and the corresponding charge polarizations are 
${\cal P}_n=\text{sgn}[\mathsf{P}(\boldsymbol{\varrho}_n)]$
with
\begin{equation}
\mathsf{P}(\mathbf{k})=\int\mathsf{J}_{\mathbf{h}_\text{so}}(\mathbf{k})d{k_x}d{k_y}=\int\partial_{k_x}g_3(k_x)\partial_{k_y}g_4(k_y)dk_xdk_y=g_3(k_x)g_4(k_y).
\end{equation}
It is clear that the second-order topology of the Hamiltonian~\eqref{sep} can be characterized by $W=\sum_n{\cal C}_{n}{\cal P}_{n}/4$.

\subsection{2D inseparable systems}

For certain 2D inseparable systems described by \begin{equation}
{\cal H}(\mathbf{k})=h_1(\mathbf{k})\Gamma_1+h_2(\mathbf{k})\Gamma_2+h_3(\mathbf{k})\Gamma_3+h_4(\mathbf{k})\Gamma_4,
\label{2DHamiltonian}
\end{equation}we cannot directly deform that as the separable forms. Nevertheless, the lattice symmetries such as mirror or mirror-rotation  symmetry can guarantee that the second-order topology of the inseparable ${\cal H}(\mathbf{k})$ is same as the 1D Hamiltonian in the high-symmetric lines, of which this 1D Hamiltonian is separable. Therefore, our proposal of $W$ is also applicable for this inseparable ${\cal H}(\mathbf{k})$. Considering a inseparable Hamiltonian ${\cal H}(\mathbf{k})$ which has mirror-rotation symmetry $M_{xy}$, one can express ${\cal H}(\mathbf{k})$ on the high-symmetry line $k_x=k_y=k$ as
\begin{equation}
{\cal H}(k)=h_1(k)\Gamma_1+h_2(k)\Gamma_2+h_3(k)\Gamma_3+h_4(k)\Gamma_4
\label{Mirror}
\end{equation}
without changing the topology, which can be further written as \begin{equation}
{\cal H}'(k)=H_+(k)\oplus H_-(k) 
\end{equation}
by a unitary operator ${\cal H}'(k)=U^{-1}{\cal H}(k)U$. Here $H_\pm(k)$ acts on the mirror-rotation subspace. The topological invariant $W$ defined by ${\cal H}(\mathbf{k})$ and ${\cal H}'(k)$ is equivalent to the second-order topology of ${\cal H}(k)$. It should be noted that this inseparable system protected by $M_{xy}$ only has the topological phase transition induced by bulk energy gap closing. We have shown these results in the second numerical example of the main text, where this 2D Hamiltonian can be firstly deformed to a form which is similar to the Hamiltonian~\eqref{insep}. And then, $M_{xy}$ can drive it to the separable Hamiltonian~\eqref{Mirror} with the same topology, which can be characterized by $W$. 

\begin{figure*}[bp]
\centering
\includegraphics[width=1.0\textwidth]{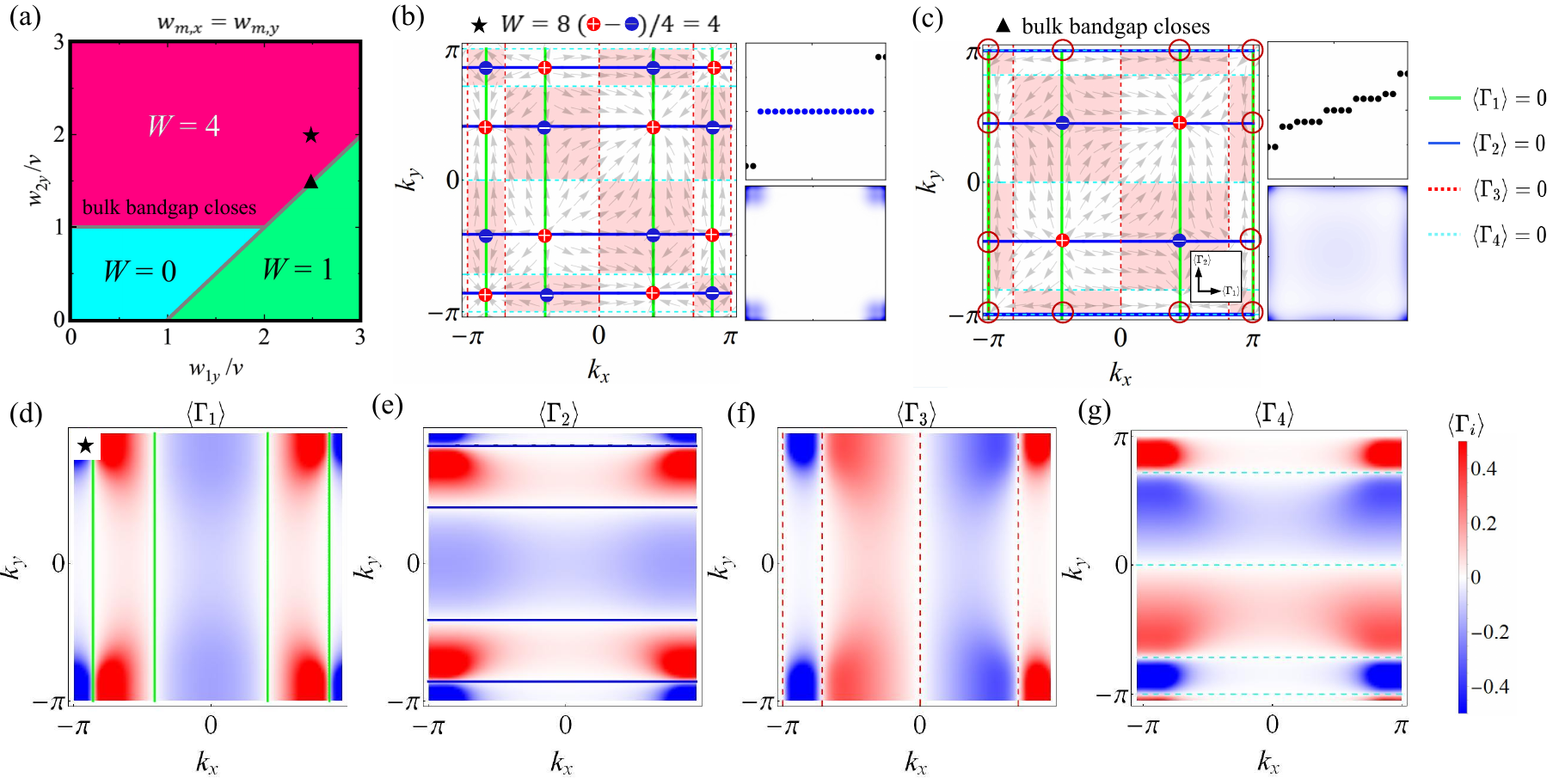}
\caption{(a) Phase diagrams of the second-order topological phase indicated by $W$. (b)-(c) Pseudospin structures of $\boldsymbol{\Theta}(\mathbf{k})$, giving $W=4$ at $w_{2,y}/v=2.0$ (star), bulk band gap closing induced topological transitions at $w_{2,y}/v=1.5$ (triangle), respectively. The insets show the zero-energy states at OBCs and their real-space distributions. (d)-(f) Pseudospin polarizations of $\langle\Gamma_{i}(\mathbf{k})\rangle$ with $i=1,2,3,4$, where $\mathbf{h}_\text{m}=0$ and  $\mathbf{h}_\text{so}=0$ are given by $\langle\Gamma_{1,2}(\mathbf{k})\rangle$ and $\langle\Gamma_{3,4}(\mathbf{k})\rangle$, respectively. Here the other parameters are $w_{1(2),x}=w_{1(2),y}$ and $w_{1,y}/v=2.5$.}
\label{fig:s1}
\end{figure*}

\subsection{Generalization in $d$D $d$th-order topological systems}
We can extend these $2$D results to the $d$D $d$th-order topological phases. For the $d$D lattice systems hosting $\mathbb{Z}$-classified $d$th-order topological phases, the corresponding momentum-space Hamiltonian reads
\begin{equation}
{\cal H}(\mathbf{k})=\sum^{2d}_{i=1}h_i(\mathbf{k})\Gamma_i.
\label{dDHamiltonian}
\end{equation}
We separate $\mathbf{h}$-vector into two parts, i.e., $\mathbf{h}=(\mathbf{h}_\text{m},\mathbf{h}_\text{so})$, where $\mathbf{h}_\text{m}=(h_1,\cdots,h_d)$ describes the band dispersion and $\mathbf{h}_\text{so}=(h_{d+1},\cdots,h_{2d})$ denotes the SO coupling. 
The $n$th nodal point $\boldsymbol{\varrho}_n$ are determined by $\mathbf{h}_\text{m}(\boldsymbol{\varrho}_n)=0$. Hence the polarized topological charges ${\cal C}_n{\cal P}_n$ are given by 
\begin{equation}
{\cal C}_n=\text{sgn}[\mathsf{J}_{\mathbf{h}_\text{m}}(\boldsymbol{\varrho}_n)],~~\mathsf{J}_{\mathbf{h}_\text{m}}(\mathbf{k})=\begin{vmatrix}
\partial_{k_1}h_1(\mathbf{k}) &\cdots & \partial_{k_d}h_1(\mathbf{k})\\
\vdots & \ddots & \vdots\\
\partial_{k_1}h_d(\mathbf{k}) & \cdots& \partial_{k_d}h_d(\mathbf{k})
\end{vmatrix}
\end{equation}
and
\begin{equation}
{\cal P}_n=\text{sgn}[\mathsf{P}(\boldsymbol{\varrho}_n)],~~\mathsf{P}(\mathbf{k})=\int \mathsf{J}_{\mathbf{h}_\text{so}}(\mathbf{k})d\mathbf{k}.
\end{equation}
For the typical case that $h_{i}(\mathbf{k})$ is the function of $k_{d+i}$ with $i=1,2,\cdots,d$, i.e., $h_{d+i}(\mathbf{k})=h_{d+i}(k_{i})$, we have 
\begin{equation}
\mathsf{P}(\mathbf{k})=h_{d+1}(\mathbf{k})\cdots h_{2d}(\mathbf{k}).
\end{equation} 
The above results always work for the $d$D separable and the certain inseparable Hamiltonian similar to the above 2D cases. Hence the $d$D $d$th-order topological phases are finally determined by $1/(-2)^d$ of the total polarized topological charges in the BZ, i.e.,
\begin{align}
W=\frac{1}{(-2)^d}\sum_n{\cal C}_n{\cal P}_n,
\label{seq:dDcharacterization}
\end{align}
which provides an unified momentum-space characterization for the higher-order topological phases.

To prove the above results, we rotate the special separable system to the generic one. Let us first consider a special $d$D separable Hamiltonian
\begin{equation}
{\cal H}(k_1,k_2,\cdots, k_d)=\sum^d_{i=1}\left[g_i(k_i)\tilde{\Gamma}_i+g_{i+d}(k_i)\tilde{\Gamma}_{i+d}\right],
\label{insep_g}
\end{equation}
with $\{\tilde{\Gamma}_i,\tilde{\Gamma}_j\}=\delta_{ij}$. For an even-dimensional system (i.e., $d$ takes an even), we deform ${\cal H}(k_1,k_2,\cdots, k_d)$ without changing the topology by choosing
\begin{equation}
\begin{split}
h_\alpha=\theta g_\alpha(k_\beta)+\phi g_{\alpha+1}(k_{\beta+1}),~h_{\alpha+1}=\mu g_\alpha(k_\beta)+\nu g_{\alpha+1}(k_{\beta+1}),~
\Gamma_\alpha=\theta\tilde{\Gamma}_\alpha+\mu\tilde{\Gamma}_{\alpha+1},~\Gamma_{\alpha+1}=\phi \tilde{\Gamma}_\alpha+\nu\tilde{\Gamma}_{\alpha+1},
\end{split}
\label{hg_1}
\end{equation}
for $\alpha=i$ and $\beta=i$ giving $\mathbf{h}_\text{m}$ and $\alpha=d+i$ and $\beta=i$ giving $\mathbf{h}_\text{so}$, where $i=1,3,5,\cdots,d-1$. Here $\theta$, $\phi$, $\mu$, and $\nu$ are real number. We finally obtain the generic $d$D separable Hamiltonian as follows:
\begin{equation}
\begin{split}
{\cal H}(k_1,k_2,\cdots, k_d)=\sum^d_{i=1}\left[h_i(\mathbf{k})\Gamma_i+h_{d+i}(\mathbf{k})\Gamma_{d+i}\right],
\label{sep_g1}
\end{split}
\end{equation}
where $\theta$, $\phi$, $\mu$, and $\nu$ should obey
\begin{equation}
\theta^2+\mu^2=1,~~\phi^2+\nu^2=1,~~\theta\phi+\mu\nu=0,
\label{cond}
\end{equation}
so that $\{\Gamma_i,\Gamma_j\}=\delta_{ij}$ and $\{\Gamma_l,\tilde{\Gamma}_m\}=0$ are guaranteed. Hence, the topological charges are written as ${\cal C}_n=\text{sgn}[\mathsf{J}_{\mathbf{h}_\text{m}}(\boldsymbol{\varrho}_n)]$, where
\begin{equation}
\begin{split}
\mathsf{J}_{\mathbf{h}_\text{m}}(\mathbf{k})=\begin{vmatrix}
\partial_{k_1}h_1(\mathbf{k}) &\cdots & \partial_{k_d}h_1(\mathbf{k})\\
\vdots & \ddots & \vdots\\
\partial_{k_1}h_d(\mathbf{k}) & \cdots& \partial_{k_d}h_d(\mathbf{k})
\end{vmatrix}
=(\nu\theta-\mu\phi)^{d/2}\partial_{k_1}g_1(k_1)\partial_{k_2}g_{2}(k_2)\cdots\partial_{k_d}g_{d}(k_d).
\end{split}
\label{even_c}
\end{equation}
The corresponding charge polarizations are 
${\cal P}_n=\text{sgn}[\mathsf{P}(\boldsymbol{\varrho}_n)]$
with
\begin{equation}
\begin{split}
\mathsf{P}(\mathbf{k})=\int\mathsf{J}_{\mathbf{h}_\text{so}}(\mathbf{k})d{\mathbf{k}}=(\nu\theta-\mu\phi)^{d/2}g_d(k_1)g_{d+1}(k_2)\cdots g_{2d}(k_d).
\end{split}
\label{even_p}
\end{equation}
It is clear that an additional condition
\begin{equation}
(\nu\theta-\mu\phi)^d>0
\label{even_limit}
\end{equation}
is emerged to keep the sign of each polarized topological charge being unchanged. With above results, the $d$th-order topology of the Hamiltonian~\eqref{dDHamiltonian} or Hamiltonian~\eqref{sep_g1} can be characterized by $W=\sum_n{\cal C}_{n}{\cal P}_{n}/(-2)^d$.
Actually, we see that the previous $2$D proof is a special case with $\theta=\mu=\nu=1/\sqrt{2}$ and $\phi=-1/\sqrt{2}$. 

For an odd-dimensional system, we still take the forms of Eq.~\eqref{hg_1}, but there is $i=1,3,5,\cdots,d-2$. The generic $d$D separable Hamiltonian is
\begin{equation}
\begin{split}
{\cal H}(k_1,k_2,\cdots, k_d)=\sum^{d-1}_{i=1}\left[h_i(\mathbf{k})\Gamma_i+h_{d+i}(\mathbf{k})\Gamma_{d+i}\right]+h_d(\mathbf{k})\tilde{\Gamma}_d+h_{2d}(\mathbf{k})\tilde{\Gamma}_{2d},
\label{sep_g2}
\end{split}
\end{equation}
with $h_d(\mathbf{k})=g_d(k_d)$ and $h_{2d}(\mathbf{k})=g_{2d}(k_d)$. The general results are similar to the odd-dimensional system, where $\theta$, $\phi$, $\mu$, and $\nu$ still need to satisfy Eq.~\eqref{cond}. Only the parameters in Eqs.\eqref{even_c} and \eqref{even_p} are changed as $(\nu\theta-\mu\phi)^{(d-1)/2}$. And then, the corresponding limited condition is written as $(\nu\theta-\mu\phi)^{d-1}>0$.

\begin{figure*}[tbp]
\centering
\includegraphics[width=1.0\textwidth]{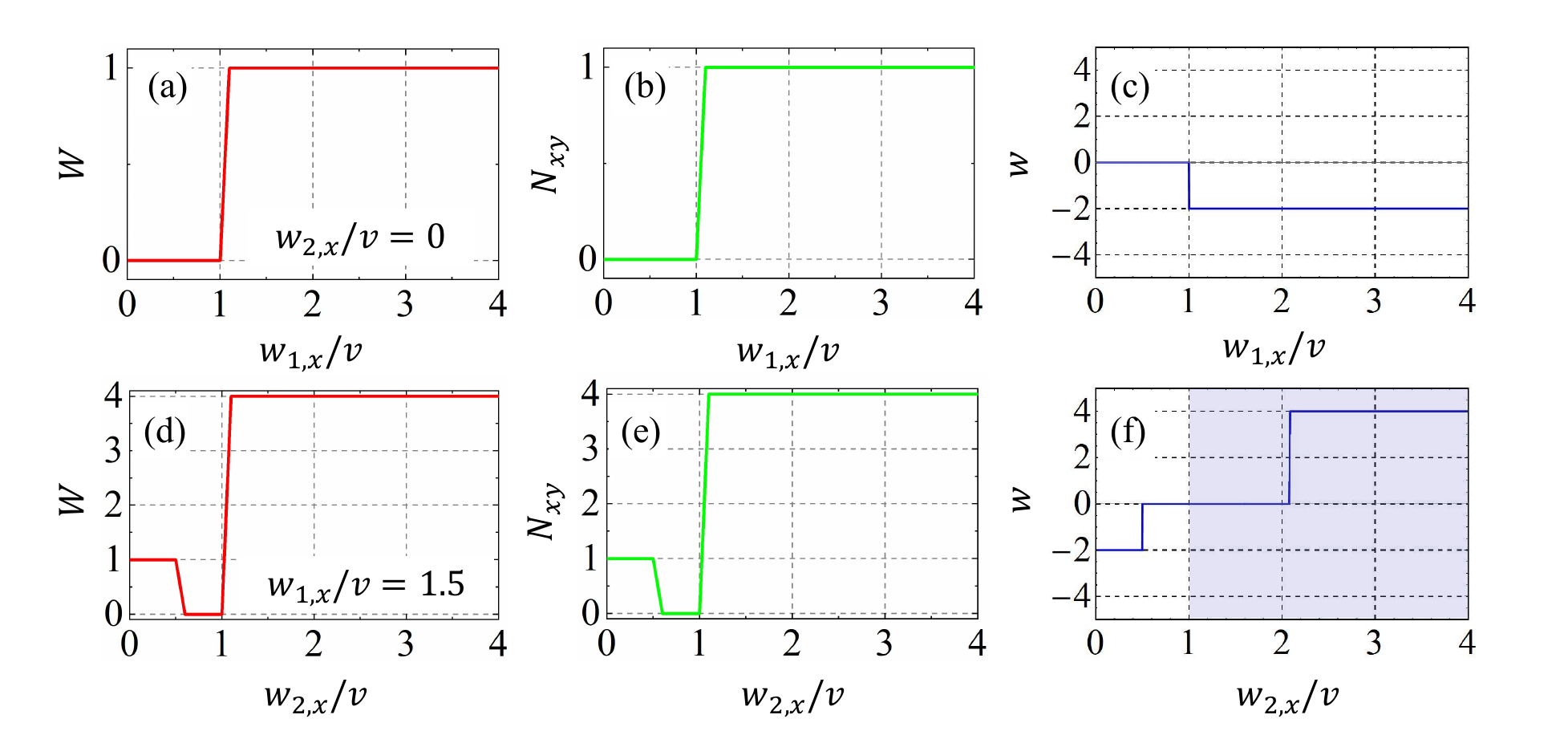}
\caption{The numerical results with the polarized topological charge ($W$), multipole chiral number ($N_{xy}$), and mirror-graded winding number ($w$). The parameters are $w_{2,x}/v=0$ for (a), (b), and (c), and $w_{1,x}/v=1.5$ for (d), (e), and (f).  $W$ and $N_{xy}$ can exactly characterize the second-order topology, but $w$ is invalid for the large-number corner states when $w_{2,x}/v>1.0$. Here the system size is $40\times 40$ for calculating $N_{xy}$.}
\label{fig:s2}
\end{figure*}

\section{Numerical results}\label{nul}

\subsection{Numerical results of 2D extend BBH model}

We consider a 2D extended BBH model with the separable Hamiltonian~\cite{benalcazar2022chiral-s}. The $h$-components are written as
\begin{equation}
\begin{split}
&h_{1}=v+ w_{1,x}\cos k_{x}+ w_{2,x}\cos 2k_{x},\\
&h_{2}=v+ w_{1,y}\cos k_{y}+ w_{2,y}\cos 2k_{y},\\
&h_{3}=w_{1,x}\sin k_{x}+w_{2,x}\sin 2k_{x},\\
&h_{4}=w_{1,y}\sin k_{y}+w_{2,y}\sin 2k_{y}.\\
\end{split}
\end{equation}
The $\boldsymbol{\Gamma}$ matrices are $\Gamma_1=-\tau_x\sigma_z$, $\Gamma_2=\tau_x\sigma_x$, $\Gamma_3=-\tau_y\sigma_0$, and $\Gamma_4=-\tau_x\sigma_y$. This system only possesses chiral symmetry under a chiral operator $\Gamma_5=\tau_z\sigma_0$, and thus belongs to the class AIII. 

When we consider the parameter of $w_{m,x}=w_{m,y}$ with $m=1,2$, the phase diagram indicated by $W$ is given in Fig.~\ref{fig:s1}(a). It is seen that the case of $W=4$ is captured by sixteen polarized topological charges in Fig.~\ref{fig:s1}(b), where the pseudospin polarizations of $\langle\Gamma_{i}(\mathbf{k})\rangle$ with $i=1,2,3,4$ are shown in Figs.~\ref{fig:s1}(d)-\ref{fig:s1}(f). Thus we can identify the values of topological charges by $\boldsymbol{\Theta}(\mathbf{k})$. However, a zero-value polarized topological charge emerges at the momentum point $(-\pi,-\pi)$ when $w_{2,y}=1.5v$ [see Fig.~\ref{fig:s1}(c)], which implies that the system occurs a topological phase transition. Particularly, this topological charge is merged at the nodal point of $\mathbf{h}_\text{so}$, i.e., $\mathbf{h}_\text{so}=0$, giving a bulk energy gap closing. 

In addition, this system also has mirror, inversion, and $C_4$-rotation symmetries with $M_{x}=\tau_x$, $M_{y}=\tau_y\sigma_y$, $I=\tau_z\sigma_y$, and $C_4=\left[(\tau_x+i\tau_y) \sigma_z-(\tau_x-i\tau_y) \sigma_x\right]/2$ for $w_{m,x}=w_{m,y}$. This induces a twofold mirror-rotation symmetry $M_{xy}H(k_x,k_y)M_{xy}^{-1}=H(k_y,k_x)$ with $M_{xy}=C_4M_y$ Following Ref.~\cite{liu2019second-s}, a single zero mode in each corner can be characterized by a mirror-graded winding number $w=w_+-w_-$ with
\begin{equation}
w_\pm=\int _\text{BZ}\frac{\text{d}k}{4\pi i}\text{Tr}\left[S'H^{-1}_{\pm}(k)\frac{\text{d}H_{\pm}(k)}{\text{d}k}\right].
\end{equation} 
Here we have $k_x=k_y=k$ and $H_{\pm}(k)=2h_1\sigma_x\pm 2h_3\sigma_y$ with $S'=\sigma_z$. We show the numerical results of the different topological index in Fig.~\ref{fig:s2}. It is seen that the polarized topological charge ($W$) and multipole chiral number ($N_{xy}$) can exactly characterize the second-order topology in the momentum-space and real-space, respectively. However, $w$ is invalid for the large-number corner states, as shown in Fig.~\ref{fig:s2}(f). 

\begin{figure*}[bp]
\centering
\includegraphics[width=1.0\textwidth]{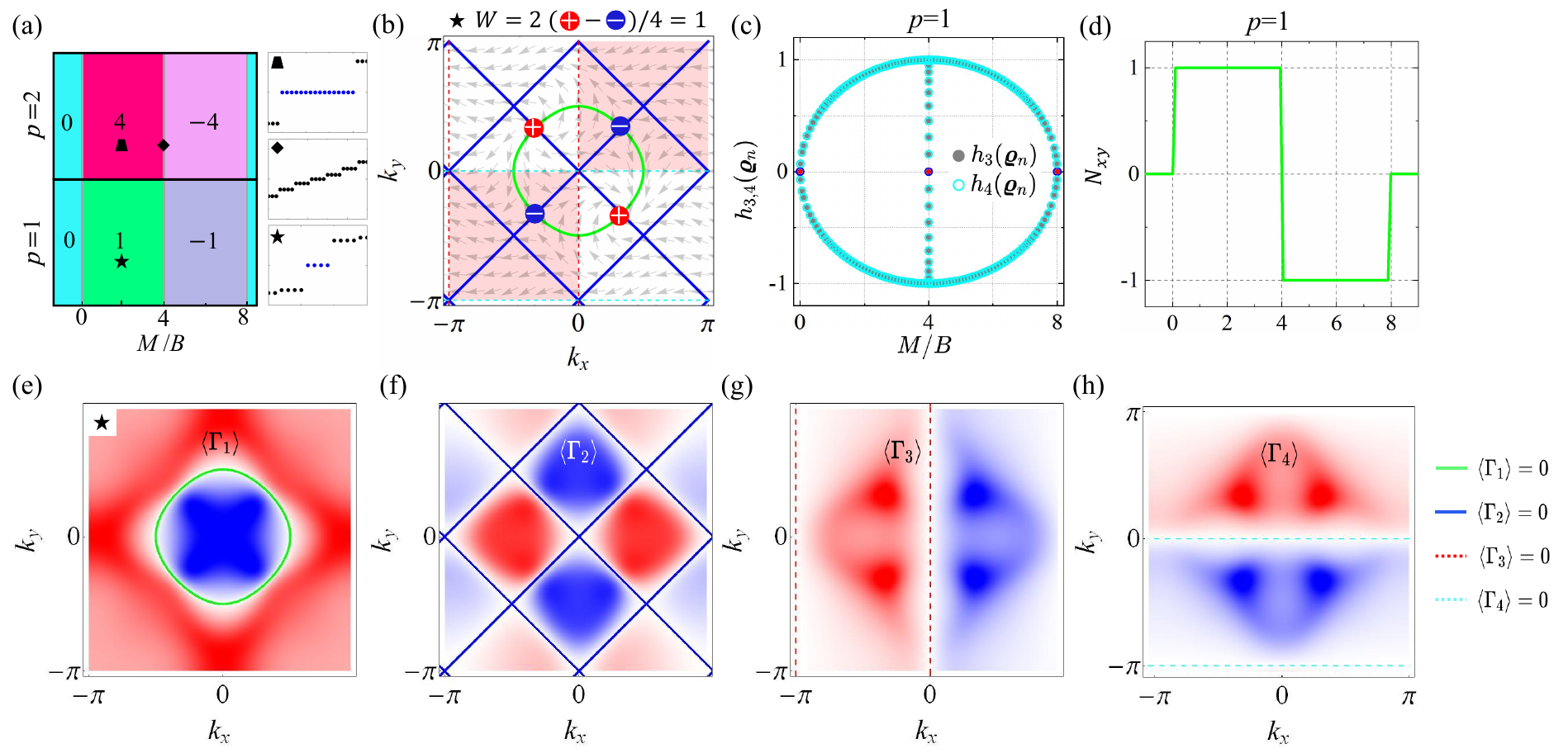}
\caption{(a) Phase diagrams of the second-order topological phases of the class BDI. The zero energy states of OBCs are given at the parameters $(p,M/B)=(2,2.0)$ (trapezoid), $(2,4.0)$ (rhombus), and $(1,2.0)$ (star), respectively. (b) Pseudospin structures of $\boldsymbol{\Theta}(\mathbf{k})$ capture four polarized topological charges, giving $W=1$. (c) The numerical results of $h_{3,4}(\varrho_n)=0$, giving ${\cal P}_n=0$ at $M/B=0$, $4$, and $8$ and showing a topological phase transition induced by bulk energy gap closing. (d) The numerical results for $p=1$ by calculating the multipole chiral number $N_{xy}$ with the system size $40\times 40$. (e)-(h) Pseudospin polarizations of $\langle\Gamma_{i}(\mathbf{k})\rangle$ with $i=1,2,3,4$, where $\mathbf{h}_\text{m}=0$ and  $\mathbf{h}_\text{so}=0$ are given by $\langle\Gamma_{1,2}(\mathbf{k})\rangle$ and $\langle\Gamma_{3,4}(\mathbf{k})\rangle$, respectively. }
\label{fig:s3}
\end{figure*}

\subsection{Numerical results of 2D extend BHZ model}

We further consider a 2D extend Bernevig-Hughes-Zhang (BHZ) model with the inseparable Hamiltonian. The $h$-components read
\begin{equation}
\begin{split}
&h_1=M-2B(2-\cos pk_x-\cos pk_y),~h_3=\sin pk_x,\\
&h_2=(\cos 2pk_x -\cos 2pk_y),~h_4=-\sin pk_y,\\
\end{split}
\label{eq:s-EBHZ}
\end{equation}
with an integer $p$. The $\boldsymbol{\Gamma}$ matrices are given by $\Gamma_1=\sigma_z$, $\Gamma_2=\tau_x\sigma_x$, $\Gamma_3=\tau_z\sigma_x$, and $\Gamma_4=\sigma_y$, with a chiral operator $\Gamma_5=\tau_y\sigma_x$. This Hamiltonian has time-reversal and particle-hole symmetries with ${\cal T}=\tau_x {\cal K}$ and ${\cal P}=\tau_z\sigma_x{\cal K}$, where $\cal K$ is complex conjugate operator. Therefore, this system belongs to the class BDI. The system also hosts $M_x=\tau_y\sigma_z$, $M_y=\tau_x$, $I=\tau_z\sigma_z$, $C_4=\text{diag}(i,1,-1,-i)$, and $M_{xy}=C_4M_y$. 

We show phase diagrams of $W$ in Fig.~\ref{fig:s3}(a). For the topological phases with $W=1$, it is seen that the structures of $\boldsymbol{\Theta}(\mathbf{k})$ figure out four polarized topological charges, as shown in Fig.~\ref{fig:s3}(b), of which $\boldsymbol{\Theta}(\mathbf{k})$ is determined by the pseudospin polarizations $\langle\Gamma_{i}(\mathbf{k})\rangle$ with $i=1,2,3,4$ in Figs.~\ref{fig:s3}(e)-\ref{fig:s3}(f). Besides, the second-order topological transitions are captured by closing bulk energy gap at the parameters $M/B=0$, $4$, and $8$, which are determined by $h_{3,4}(\varrho_n)=0$ in Fig.~\ref{fig:s3}(c) and gives ${\cal P}_n=0$. This implies the value of each polarized topological charge being zero. Besides, we also show the numerical result of $N_{xy}$ in Fig.~\ref{fig:s3}(d), which is consistent with the polarized topological charges. Hence our characterization theory provides an
elegant and unified scheme to identify the higher-order topological states in the momentum space.

\subsection{Numerical results of 3D BBH model}

We finally consider a simple 3D BBH model with the separable Hamiltonian ${\cal H}(\mathbf{k})=\sum^6_{i=1} h_i(\mathbf{k})\Gamma_i$~\cite{benalcazar2017quantized,benalcazar2017electric-s}. The $h$-components read
\begin{equation}
\begin{split}
&h_1=\gamma+\lambda \cos k_x,~h_2=\gamma+\lambda \cos k_z,~h_3=\gamma+\lambda \cos k_y,\\
&h_4=\lambda \sin k_x,~~~h_5=\lambda \sin k_y,~~~h_6=\lambda \sin k_z.\\
\end{split}
\label{eq:3DBBH}
\end{equation}
The $\boldsymbol{\Gamma}$ matrices are given by $\Gamma_1=\rho_x$, $\Gamma_2=\rho_z\tau_x$, $\Gamma_3=-\rho_z\tau_y\sigma_y$, $\Gamma_4=-\rho_z\tau_y\sigma_z$, $\Gamma_5=-\rho_z\tau_y\sigma_x$, and $\Gamma_6=\rho_y$, with a chiral operator $\Gamma_7=\rho_z\tau_z$.
Here these $\boldsymbol{\Gamma}$ matrices obey $\text{Tr}[\Gamma_7\Gamma_1\Gamma_2\Gamma_3\Gamma_4\Gamma_5\Gamma_6]=8i$. It is well-known that there is a quantized octupole moment $o_{xyz}=1/2$ when $|\lambda|>|\gamma|$~\cite{benalcazar2017quantized-s,benalcazar2017electric-s}, which characterizes that this 3D system can host eight corner states and there is only one zero mode in each corner. Besides, the multipole chiral number gives $N_{xyz}=1$ to characterize these topological states~\cite{benalcazar2022chiral-s}. For this case of $|\lambda|>|\gamma|$, we can obtain eight topological charges 
\begin{equation}
{\cal C}_n=\text{sgn}[-\lambda^3\sin \varrho_l \sin \varrho_m \sin \varrho_s]
\end{equation}
and their polarization
\begin{equation}
{\cal P}_n=\text{sgn}[\lambda^3\sin \varrho_l \sin \varrho_m \sin \varrho_s],
\end{equation}
where $\boldsymbol{\varrho}_n=(\varrho_l,\varrho_m,\varrho_s)$ with $h_1(\varrho_l )=h_2(\varrho_m)=h_3(\varrho_s)=0$ and $n=1,2,\cdots,8$. Here $\varrho_{l,m,s}=\pm\arccos(-\gamma/\lambda)$. Hence each polarized topological charge is given by 
\begin{equation}
{\cal C}_n{\cal P}_n=-\text{sgn}[\lambda^6\sin^2 \varrho_l \sin^2 \varrho_m \sin^2 \varrho_s]=-1
\end{equation}
and we have 
\begin{equation}
W=-\frac{1}{8}\sum_n {\cal C}_n{\cal P}_n=1.
\end{equation}
However, there is no topological charge for $|\lambda|<|\gamma|$, where the system is the third-order topologically trivial. 
For $|\lambda|=|\gamma|$, we have $\varrho_{l,m,s}=0$ or $-\pi$, giving the topological phase transition where the bulk gap is closing.
It is clear that our result $W$ is consistent with $o_{xyz}$ and $N_{xyz}$. Compared with $o_{xyz}$ and $N_{xyz}$, this $W$ is easily measured in the realistic quantum simulation experiments.

\section{Higher-order topological phase transitions}\label{hotpt}

For the higher-order topological phases, it is well-known that the topological properties not only depend on the bulk topology, but can also be related to the boundary properties of the system. The main reason is that the high-dimensional boundary of the system has an open energy gap, which means that the emergence of topological phase transition can not only close the bulk energy gap but can close the boundary energy gap of different dimensions. Based on our topological characterization theory, these two types of higher-order topological transitions can be identified by the emergence of zero-polarized topological charges. Specifically, the topological charge moves to the nodal lines or surfaces of the SO field $\mathbf{h}_\text{so}$, i.e., there are $N_j$ zero-value $h_j$-components with $j\in \{d+1,\cdots,2d\}$ and $0<N_j<d$, giving a topological transition induced by boundary band gap closing. However, the topological charge moves to the nodal points of $\mathbf{h}_\text{so}$, i.e., $\mathbf{h}_\text{so}=0$, giving a topological transition induced by bulk band gap closing.

For a $2$D second-order topological phase described by ${\cal H}(\mathbf{k})=\sum^{4}_{i=1}h_i(\mathbf{k})\Gamma_i$, the topological transitions induced by bulk (edge) band gap closing can be captured by the topological charges moving to the nodal points (lines) of $\mathbf{h}_m=(h_1,h_2)$, giving $h_{1,2,3,4}=0$ ($h_{1,3,4}=0$ for $x$ edge or $h_{2,3,4}=0$ for $y$ edge), as shown in Figs.~2(c),~2(b), and S1(c). For a 3D third-order topological phases described by ${\cal H}(\mathbf{k})=\sum^6_{i=1} h_i(\mathbf{k})\Gamma_i$, the third-order topological phase transitions are naturally identified by the zero-value polarized topological charges, giving the bulk band gap closing with $h_{1,2,3,4,5,6}=0$, or surface band gap closing with $h_{1,2,3,4,5}=0$ ($xy$ surface), $h_{1,2,3,4,6}=0$ ($xz$ surface), and $h_{1,2,3,5,6}=0$ ($yz$ surface), or hinge band gap closing with $h_{1,2,3,4}=0$ ($x$ hinge), $h_{1,2,3,5}=0$ ($y$ hinge), and $h_{1,2,3,6}=0$ ($z$ hinge). Hence our topological characterization provides an intuitive picture to identify the higher-order topological transitions in the momentum space. 

\section{Experimental scheme in $^{87}$Rb atomic system}\label{exp}

We provide an experimental scheme to realize the chiral-symmetric second-order topological phases in $^{87}$Rb cold atomic system. This scheme is based on the realization of BHZ model in the recent cold atom system~\cite{lv2021measurement-s}. By employing four atomic hyperfine levels $\ket{a}=\ket{2,-1}$, $\ket{b}=\ket{1,-1}$, $\ket{c}=\ket{2,0}$, and $\ket{d}=\ket{1,0}$, we use the microwaves to couple $\{\ket{a},\ket{b}\}$, $\{\ket{c},\ket{d}\}$, $\{\ket{a},\ket{d}\}$, and $\{\ket{c},\ket{b}\}$ with Rabi frequencies $\Omega_1$, $\Omega_2$, $\Omega_3$, and $\Omega_4$. By using the bare-state basis $\{\ket{a},\ket{b},\ket{c},\ket{d}\}$, the interacting Hamiltonian is given by
\begin{equation}
\begin{split}
H=&(\omega_{a}-\omega_{b})\ket{a}\bra{a}+(\omega_{c}-\omega_{b})\ket{c}\bra{c}+(\omega_{d}-\omega_{b})\ket{d}\bra{d}\\
&+({\Omega_1}e^{i\omega_1 t} e^{i\varphi_1}\ket{a}\bra{b}+{\Omega_2}e^{i\omega_2 t} e^{i\varphi_2}\ket{c}\bra{d}\\
&+{\Omega_3}e^{i\omega_3 t} e^{i\varphi_3}\ket{a}\bra{d}+{\Omega_4}e^{i\omega_4 t} e^{i\varphi_4}\ket{c}\bra{b}+H.c. )\\
\end{split}
\label{EffHamiltonian}
\end{equation}
where $\omega_{j}$ are the energy frequencies of  $\ket{j}$ with $j=a,b,c,d$.
Here $\Omega_s$, $\omega_s$, and $\varphi_s$ with $s=1,2,3,4$ are the Rabi frequencies, frequencies, and phase of the controlling microwaves, respectively. By turning the Hamiltonian to the reference frame $U=e^{-i\omega_1t}\ket{a}\bra{a}+\ket{b}\bra{b}+e^{-i\omega_4t}\ket{c}\bra{c}+e^{-i(\omega_1-\omega_3)t}\ket{d}\bra{d}$, we can obtain the effective Hamiltonian $H_\text{eff}=i\partial_tU^\dagger U + U^\dagger H U$ as
\begin{equation}
H_\text{eff}=
\begin{pmatrix}
-\Delta_1 & {\Omega_1}e^{-i\varphi_1}& 0 & {\Omega_3}e^{-i\varphi_3}\\
{\Omega_1}e^{i\varphi_1} & 0 & {\Omega_4}e^{i\varphi_4} & 0\\
0 & {\Omega_4}e^{-i\varphi_4} &-\Delta_4 & {\Omega_2}e^{-i\Delta^\prime t} e^{-i\varphi_2}\\
{\Omega_3}e^{i\varphi_3}&0&{\Omega_2}e^{i\Delta^\prime t} e^{i\varphi_2}&\Delta_3-\Delta_1\\
\end{pmatrix},
\end{equation}
where $\Delta_1=\omega_1-(\omega_a-\omega_b)$, $\Delta_2=\omega_2-(\omega_c-\omega_d)$, $\Delta_3=\omega_3-(\omega_a-\omega_d)$, $\Delta_4=\omega_4-(\omega_c-\omega_b)$, and $\Delta^\prime =\omega_1+\omega_2-\omega_3-\omega_4=\Delta_1+\Delta_2-\Delta_3-\Delta_4$, respectively. In order to arrive at the second-order topological phases of the class BDI, we choose $\Delta^\prime=0$, i.e., $\Delta_1+\Delta_2=\Delta_3+\Delta_4$, which induce a time-independent Hamiltonian. We further tune $\Delta_1=\Delta_2=\Delta_3=\Delta_4=2\Delta$ by sweeping $\omega_{1,2,3,4}$ and shift the energy levels by $\Delta$ to obtain
\begin{equation}
H_\text{eff}=
\begin{pmatrix}
-\Delta & {\Omega_1}e^{-i\varphi_1}& 0 & {\Omega_3}e^{-i\varphi_3}\\
{\Omega_1}e^{i\varphi_1} & \Delta & {\Omega_4}e^{i\varphi_4} & 0\\
0 & {\Omega_4}e^{-i\varphi_4} &-\Delta& {\Omega_2}e^{-i\varphi_2}\\
{\Omega_3}e^{i\varphi_3}&0&{\Omega_2}e^{i\varphi_2}&\Delta\\
\end{pmatrix}.
\end{equation}
In realistic experiments, we have 
\begin{equation}
\Omega_1=-\Omega_2=\sqrt{h_3^2+h_4^2},~~
\Omega_3=\Omega_4=h_2,~~
\Delta=-h_1,~~\varphi_1=-\varphi_2=\arctan (h_4/h_3), ~~
\varphi_3=\varphi_4=0.
\end{equation}
The effective Hamiltonian in experimental conditions can be arrived at
\begin{equation}
H_\text{eff}=
\begin{pmatrix}
h_1 & h_3-ih_4& 0 & h_2\\
h_3+ih_4& -h_1& h_2 & 0\\
0 & h_2 & h_1 & -h_3-ih_4\\
h_2&0&-h_3+ih_4&-h_1\\
\end{pmatrix}
=h_1\sigma_z+h_2\tau_x\sigma_x+h_3\tau_z\sigma_x+h_4\sigma_y.
\end{equation}
Compared with the experiment of realizing BHZ model, it is seen that the $\mathbf{k}$-dependent $h_2$ term opens band gap of the helical states and induces the corner states, rendering the second-order topological phases. 

\bibliographystyle{apsrev4-1}

\end{document}